\documentclass[journal,12pt,draftcls,onecolumn,a4paper]{IEEEtran}
\usepackage{cite}
\usepackage{amsmath,amssymb,amsfonts}
\usepackage{algorithmic}
\usepackage{graphicx}
\usepackage{tikz}
\usepackage{subcaption}
\usepackage{changepage}
\allowdisplaybreaks 
\usetikzlibrary{shapes,arrows}
\usepackage{multicol,lipsum}
\usepackage{float}
\usepackage{lipsum}
\usepackage{mathtools}
\usepackage{cuted}
\usepackage{hyperref}
\usepackage{textcomp}
\usepackage{amsmath}
\usepackage{esvect}
\usepackage{bm}
\allowdisplaybreaks

\usepackage{amsmath,bbm,amssymb,amsfonts,amstext,amsopn}

\DeclareMathOperator{\Tr}{Tr}
\DeclareMathOperator{\Rank}{Rank}

\usepackage{lipsum,amsmath,multicol}
\usepackage{amsthm}
\usepackage{amssymb}
\usepackage{epstopdf}

\usepackage{amsthm}
\usepackage{cite}
\usepackage{balance}
\allowbreak
\usepackage{url}
\usepackage{epsfig}
\usepackage{setspace}
\usepackage{stmaryrd}
\usepackage{psfrag}	
\usepackage{multirow}
\usepackage{float}
\usepackage{amsthm}

\newcommand{\subparagraph}{}
\usepackage[compact]{titlesec}
\usepackage[english]{babel}
\usepackage[process=auto]{pstool}
\usepackage{amsthm}
\theoremstyle{remark}
\newtheorem{remark}{Remark}
\usepackage{etoolbox}
\usepackage{algorithm}
\usepackage[process=auto]{pstool}
\usepackage{epsfig}
\usepackage{caption}
\usepackage{algorithmic}
\usepackage{xcolor}

\newcommand{\jj}{{j}}

\newcommand{\e}{{e}}

\def\by{\mathbf{y}}
\def\bz{\mathbf{z}}

\def\bPsi{\boldsymbol{\Psi}}

\begin{document}
\title{Optimization-based Phase-shift Codebook Design for Large IRSs} 
\author{ Walid R. Ghanem$^{*}$,
Vahid Jamali$^{*}$,  Malte Schellmann$^{\dagger}$, Hanwen Cao$^{\dagger}$, Joseph Eichinger$^{\dagger}$, and Robert Schober$^{*}$  \\
$^{*}$Friedrich-Alexander-University Erlangen-Nuremberg, Erlangen, Germany\\ $^{\dagger}$Huawei Technologies German Research Center, Munich, Germany \quad} \vspace{-0.5cm}
\maketitle
\begin{abstract}
In this paper, we focus on large intelligent reflecting
surfaces (IRSs) and propose a new codebook construction method to obtain a set of pre-designed phase-shift configurations for the IRS unit cells. Since the complexity of online optimization and the overhead for channel estimation scale with the size of the phase-shift codebook, the design of small codebooks is of high importance. We consider both continuous and discrete phase-shift designs and formulate the codebook construction as optimization problems. To solve the optimization problems, we propose an optimal algorithm for the discrete phase-shift design and a low-complexity sub-optimal solution for the continuous design. Simulation results show that the proposed algorithms facilitate the construction of codebooks of different sizes and with different beamwidths. Moreover, the performance of the discrete phase-shift design with $2$-bit quantization is shown to approach that of the continuous phase-shift design. Finally, our simulation results show that the proposed designs enable large transmit power savings compared to the existing linear and quadratic codebook designs \cite{marirsj,quad}. 
\end{abstract}
\section{Introduction} 
Recently, intelligent reflecting surfaces (IRSs) have been proposed to shape wireless communication channels \cite{Quirs1,enerirs,marco}. \color{black} IRSs typically comprise large numbers of
reconfigurable elements and are deployed between the base station (BS) and
the user to establish a virtual line-of-sight (LoS) link. By properly configuring the elements, the IRS
can provide a high passive beamforming gain.  However, due to the required large number of IRS elements, online optimization and channel estimation in IRS-assisted systems is challenging. 

Most existing works \cite{Quirs1,enerirs,marco,irsdiscrete,irsswipt,alexpower} aim to optimize all IRS elements individually for real-time communication, which leads to computationally complex resource allocation problems for large numbers of elements \cite{marirsj}. Furthermore, it has been shown in \cite{marirsj} that for typical outdoor scenarios, hundreds or even thousands of IRS elements are required to generate a virtual LoS link, which is as strong as the direct link. Unfortunately, most algorithms in the literature, e.g., \cite{Quirs1,enerirs,marco,irsdiscrete,irsswipt,alexpower}, may not be applicable for such large IRSs since their complexity will render the optimization problem intractable. To overcome this issue, the authors of \cite{marirsj} proposed to design a set of phase-shift configurations based on a codebook framework in an offline stage, and then select the best codeword from the set in an online optimization stage for a given channel realization. Hence, the complexity of online optimization and the corresponding channel estimation overhead scale with the IRS codebook size and not explicitly with the number of reflecting elements. In \cite{irsdft,ofdmadft}, the authors employed predefined phase-shift configurations based on the discrete Fourier transform (DFT) matrix for channel estimation in IRS-assisted
systems. However, DFT-based codebooks contain the same number of phase-shift configurations as there are IRS elements, which may be too many
for practical implementation. To overcome this problem, the authors of \cite{quad} proposed a quadratic codebook design, which features a parameter to control the size of the codebook. However, the shape of the generated beams cannot be explicitly controlled. Furthermore, the authors of \cite{marirsj,irsdft,ofdmadft,quad} focus on continuous IRS phase shifts, whereas in practice, discrete IRS phase shifts may be preferred to reduce the implementation cost\cite{irsdiscrete}. To overcome this problem, recently,  the authors in \cite{codeirsdis} studied beam pattern design for discrete IRS phase shifts and proposed a corresponding sub-optimal algorithm. However, the optimal beam pattern design for discrete IRS phase shifts is still an open problem.

Motivated by the above discussion, in this paper, we propose a novel codebook design for IRS-assisted wireless communication. The codebook design is formulated in terms of two offline optimization problems for continuous and discrete IRS phase shifts, respectively. Moreover, to solve the optimization problems, we propose an optimal algorithm for the discrete phase-shift design and a low-complexity sub-optimal algorithm for the continuous phase-shift design. We show by simulation that the proposed codebook designs offer a favourable trade-off between power consumption and codebook size and attain large power savings compared to the codebook designs in \cite{marirsj,quad}. 

\textit{Notations}:  $\mathbb{C}^{N}$ represents the set of all $N \times 1$ vectors with complex valued entries. The circularly symmetric complex Gaussian distribution with mean $\mu$ and variance $\sigma^{2}$ is denoted by $\mathcal{CN}(\mu,\sigma^{2})$, $\sim$ stands for ``distributed as", and $\mathcal{E}\{\cdot\}$ denotes statistical expectation. $\nabla_{\mathbf{x}}f(\mathbf{x})$ denotes the gradient vector of function $f(\mathbf{x})$ and its elements are the partial derivatives of $f(\mathbf{x})$. $\mathbf{a}\otimes \mathbf{b}$ denotes the Kronecker product of two vectors. $\mathbf{A}^{T}$, $\mathbf{A}^{H}$, $\Rank{(\mathbf{A})}$, and $\Tr{(\mathbf{A})}$ denote transpose, Hermitian transpose, rank, and
trace of matrix $\mathbf{A}$, respectively. $\|\mathbf{A}\|_{*}$ and $\|\mathbf{A}\|_{2}$ denote the spectral
norm and nuclear norm of matrix $\mathbf{A}$, respectively.
${\boldsymbol{\lambda}}_{\textrm{max}}(\mathbf{A})$  is the eigenvector associated with the maximum eigenvalue of matrix $\mathbf{A}$. $\mathbf{x}^*$ is the optimal value of optimization
variable $\mathbf{x}$. Finally, Diag($\mathbf{a}$) denotes a diagonal matrix with the elements of
vector $\mathbf{a}$ on its main diagonal.\vspace{-0.25cm}
\section{System Model}
In this section, we present the system and signal models and the codebook concept for IRS-assisted wireless systems.
\subsection{System Model}
We consider a downlink system comprising a single-antenna BS that serves a single-antenna user. The direct link between the BS and the user is blocked. Hence, the downlink communication is assisted by a large IRS comprising $Q$ reflecting unit cells.  We consider the case where the IRS is in the far field of the BS and the user\footnote{To facilitate the support of multiple users, the IRS can be divided into tiles \cite{marirsj}. Then, using the proposed codebook, 
	each tile may serve one user (single beam) and multiple users are served by different tiles (e.g., multiple beams are created by the IRS via different tiles)\color{black}.}.  Throughout this paper, we assume that the IRS is a planar uniform array comprising $Q_{y}$ ($Q_{z}$) unit cells spaced $d_{y}$ ($d_{z}$) meter apart in $y$-direction ($z$-direction), indexed by $q_{y}=0,\dots,Q_{y}-1$ ($q_{z}=0,\dots,Q_{z}-1$), where $Q=Q_{y} Q_{z}$ and each unit cell has an area of $A_{\textrm{uc}}=d_{y} d_{z}$\color{black}. Moreover, for future reference, we define $A_{y}(\bPsi_i,\bPsi_r)=A_{y}(\bPsi_i)+A_{y}(\bPsi_r)$ and $A_{z}(\bPsi_i,\bPsi_r)=A_{z}(\bPsi_i)+A_{z}(\bPsi_r)$ with $A_{y}(\bPsi)=\sin(\theta)\sin(\phi)$ and $A_{z}(\bPsi)=\cos(\theta)$, where $\bPsi_{i} = (\theta_{i}, \phi_{i})$ and $\bPsi_{r} = (\theta_{r}, \phi_{r})$ are the
angle of arrival (AoA) of the incident wave and the angle of departure (AoD) of the reflected wave,
respectively, and $\theta$ and $\phi$ denote the elevation and azimuth angles, respectively, see \cite{marirsj} for more details. For convenience of presentation, we use either $\nu_{q_{y},q_{z}}$ or $\nu_{q}$ to refer to the phase-shift applied by the $(q_{y},q_{z})$-th or equivalently the $q$-th IRS element, respectively. Moreover, $\mathbf{w} \in \mathbb{C}^{Q}$ is a vector collecting variables $w_{q_{y},q_{z}}=e^{j\nu_{q_y,q_{z}}}, \forall q_y,q_z,$ or equivalently $w_{q}=e^{j\nu_{q}}, \forall q$.  \color{black}    

\subsection{Signal Model}
Assuming the signal arrives from AoA $\bPsi_i$ and phase shifts $w_{q_{y},q_{z}}, \forall q_{y},q_{z},$ are applied by the elements of the IRS, the IRS response function towards AoD $\bPsi_r$ is given as follows \cite{marirsj,quad}:    
\begin{align}&\label{res}
g(\bPsi_i,\bPsi_r)  
=\bar{g}\sum_{q_{y}=0}^{Q_{y}-1}\sum_{q_{z}=0}^{Q_{z}-1}
\e^{{\jj k d_{y} A_{y}(\bPsi_i,\bPsi_r)}q_{y}}
\e^{{\jj k d_{z} A_{z}(\bPsi_i,\bPsi_r)}q_{z}}
\e^{\jj \nu_{q_{y},q_{z}}}, 
\end{align}
where $k=\frac{2\pi}{\lambda}$ is the wave number, $\lambda$ is the wavelength, and $\bar{g}=\frac{4\pi A_{\text{uc}}}{\lambda^{2}}$\cite{quad}. The IRS response function determines the end-to-end path loss of the IRS-assisted virtual link \cite[Lemma 1]{marirsj}, $PL_{\text{IRS}}=|g(\bPsi_i,\bPsi_r)|^{2}P_{Lt}P_{Lr}$, where $P_{Lt}$ and  $P_{Lr}$ denote the free-space path loss from the BS to the IRS and from the IRS to the user, respectively.
\subsection{Codebook Design}
In order to maximize the power reflected
in the direction where the user is located, each unit cell of the IRS has to apply an appropriate phase-shift such
that the waves propagating from all unit cells in the direction of interest add up coherently. Moreover, a number of beams are needed to cover the entire area to be illuminated by the IRS. The set of phase-shift configurations needed to generate these beams is referred to as the \textit{codebook} and a given phase-shift configuration is a \textit{codeword}. The goal of this paper is to design suitable \textit{codewords}, such that a signal arriving from any arbitrary AoA can be reflected towards any desired AoD. A straightforward way to do this is to discretize $\bPsi_i$ into intervals and to design for each interval of $\bPsi_i$, a complete codebook for reflecting the signal towards a desired discretized $\bPsi_r$. However, as can be observed from (\ref{res}), the response function depends on $\bPsi_i$ and $\bPsi_r$ only via $A_{t}(\bPsi_i,\bPsi_r), \forall t=\{y,z\}$. Thus, it is preferable to discretize $A_{t}(\bPsi_i,\bPsi_r), \forall t=\{y,z\}$, because different $\bPsi_i$ and $\bPsi_r$ can yield the same value $A_{t}(\bPsi_i,\bPsi_r), \forall t=\{y,z\}$. In other words, the same phase-shift configuration can be used to reflect different AoAs to different AoDs. To facilitate the presentation, the IRS response function is rewritten as follows:
\begin{align}&\label{res2}
g(\beta_{y} ,\beta_{z} )=\bar{g}\mathbf{w}^{H} (\by (\beta_y) \otimes \bz(\beta_z)),
\end{align}
where $\by(\beta_{y})=[1, \dots,\e^{\jj k d_y \beta_{y} (Q_y-1)}]$, $\bz(\beta_{z})=[1, \dots,\e^{\jj k d_{z} \beta_{z} (Q_{z}-1)}]$,  $\beta_{y}  \triangleq A_{y}(\bPsi_i,\bPsi_r)$, and $\beta_{z}\triangleq A_{z}(\bPsi_i,\bPsi_r)$.   

To design a codebook, we discretize $\beta_{t}\in[-\bar{\beta}_{t}/2~ \bar{\beta}_{t}/2],\forall t=\{y,z\},$ where $\bar{\beta}_{t}=\min\{4,\lambda/d_{t}\}, \forall t,$ into $M_{t},\forall t=\{y,z\},$ intervals, respectively. Thus, the codebook contains $M=M_{y}M_{z}$ codewords and each codeword is characterized by given continuous intervals in $y$- and $z$-direction, i.e., $ \mathcal{B}_{y,m_{y}}, \forall m_{y}=\{1,\dots,M_{y}\},$ and $ \mathcal{B}_{z,m_{z}}, \forall m_{z}=\{1,\dots,M_{z}\}$. In this paper, the collection of intervals $ \mathcal{B}_{y,m_{y}}$ and $ \mathcal{B}_{z,m_{z}}$ are obtained by
uniform quantization of $[-\bar{\beta}_{t}/2~ \bar{\beta}_{t}/2],\forall t=\{y,z\}$, respectively. 

\section{Continuous Phase-Shift Design}
In this section, we design the codebook assuming the IRS phase shifts are not discretized, i.e.,  $\nu_{q_{y},q_{z}} \in [-\pi,\pi]$. 
\subsection{Problem Formulation}
Mathematically, for continuous phase shifts, the optimization problem for designing a codeword $\mathbf{w}_{m_{y},m_{z}}$ for intervals $\mathcal{B}_{y,m_{y}}, \forall m_{y} \in \{1,\dots, M_{y}\}$ and $ \mathcal{B}_{z,m_{z}}, \forall m_{z} \in \{1,\dots, M_{z}\}$, is formulated as follows:
\begin{align}\label{Op2a}
&\underset  {\mathbf{w}_{m_{y},m_{z}},\alpha}{\text{maximize}}\quad\alpha\\& \nonumber\text{s.t.} \; \mathrm{C1}: |g_{m_{y},m_{z}}(\beta_y,\beta_z)|^{2} \geq \alpha, \hspace{2cm}\nonumber \\&\hspace{2.5cm} \forall \beta_y \in \hat{\mathcal{B}}_{y,m_{y}}, \beta_z \in \hat{\mathcal{B}}_{z,m_{z}}, \nonumber \\& \;\;\quad
\mathrm{C2}:|w_{q_{y},q_{z}}|=1, \nonumber \\& \hspace{2cm} \forall q_{y}=\{0,\dots,Q_{y}-1\},q_{z}=\{0,\dots,Q_{z}-1\},\nonumber	
\end{align}
where $|g_{m_{y},m_{z}}(\beta_y,\beta_z)|^{2}= |\mathbf{w}_{m_{y},m_{z}}^{H} (\by (\beta_y)) \otimes \bz (\beta_z))|^{2}$ and $\mathbf{w}_{m_{y},m_{z}} \in \mathbb{C}^{Q}$ is the phase-shift vector containing $w_{q_{y},q_{z}}, \forall q_{y},q_{z}$.  Constraint $\mathrm{C2}$ is the unit-modulus constraint for the IRS elements to ensure that each element has unit magnitude. \color{black} Moreover, to facilitate the optimization, we discretize the continuous intervals $ \mathcal{B}_{y,m_{y}}$ and $ \mathcal{B}_{z,m_{z}}$ into $P_{y}$ and $P_{z}$ discrete points, respectively, and collect these points in the two discrete sets $ \hat{\mathcal{B}}_{y,m_{y}}$ and $ \hat{\mathcal{B}}_{z,m_{z}}$, respectively. We maximize the reflected power for the $\beta_t, \forall t \in \{y,z\}$, belonging to discrete sets $ \hat{\mathcal{B}}_{y,m_{y}}$ and $ \hat{\mathcal{B}}_{z,m_{z}}$ by maximizing $\alpha$. To obtain the complete codebook of $M$ codewords, optimization problem (\ref{Op2a}) has to be solved for all sets $ \hat{\mathcal{B}}_{y,m_{y}}, \forall m_{y}=\{1,\dots,M_{y}\},$ and $ \hat{\mathcal{B}}_{z,m_{z}}, \forall m_{z}=\{1,\dots,M_{z}\}$.

There is no systematic approach for solving general non-convex optimization problems. However, in the following, we show that problem (\ref{Op2a}) can be reformulated as a difference of convex (D.C.) functions programming problem. This reformulation allows the application of a Taylor series approximation to obtain a locally optimum solution of (\ref{Op2a}) with low-computational complexity.    
\subsection{Proposed Solution}
In the following, we transform (\ref{Op2a}) into an equivalent semi-definite programming (SDP) problem. Let us define $\mathbf{W}_{m_{y},m_{z}}=\mathbf{w}_{m_{y},m_{z}}\mathbf{w}_{m_{y},m_{z}}^{H}$ and exploit the following identity\cite{alexpower}: 
\begin{align}\label{tkb}&
|\mathbf{w}_{m_{y},m_{z}}^{H} (\by (\beta_{y}) \otimes \bz (\beta_{z}))|^{2}\\&=\Tr\big(\mathbf{W}^{H}_{m_{y},m_{z}}(\by (\beta_y) \otimes \bz (\beta_z))(\by (\beta_{y}) \otimes \bz (\beta_{z}))^{H}\big).\nonumber
\end{align}

Based on (\ref{tkb}), optimization problem (\ref{Op2a}) can be rewritten in the following equivalent SDP form:
\begin{align}\label{s1b}
&\underset  {\mathbf{W}_{m_{y},m_{z}},\alpha}{\text{maximize}}\quad\alpha\\& \nonumber\text{s.t.} \; \mathrm{C1}: G_{m_{y},m_{z}}(\beta_y,\beta_z) \geq \alpha \nonumber, \forall \beta_y \in  \hat{\mathcal{B}}_{y,m_{y}}, \beta_z \in   \hat{\mathcal{B}}_{z,m_{z}},\nonumber\\& \quad \;\;
\mathrm{C2}:\Rank(\mathbf{\mathbf{W}}_{m_{y},m_{z}})=1, \mbox {C3}: \mathbf{\mathbf{W}}_{m_{y},m_{z}} \succeq	0,\;\nonumber  \;\; \nonumber\\& \quad \;\; \mbox {C4}:\text{Diag}(\mathbf{\mathbf{W}}_{m_{y},m_{z}})=\mathbf{1}_{Q},\nonumber
\end{align}
where \color{black} $G_{m_{y},m_{z}}(\beta_y,\beta_z)=\Tr\big(\mathbf{W}^{H}_{m_{y},m_{z}}(\by (\beta_y) \otimes \bz (\beta_z))(\by (\beta_y) \otimes \bz (\beta_z))^{H}\big)$ and constraints $\mathrm{C2,C3},$ and $\mathrm{C4}$ are imposed to guarantee that $\mathbf{W}_{m_{y},m_{z}}=\mathbf{w}_{m_{y},m_{z}}\mathbf{w}_{m_{y},m_{z}}^{H}$ and $|w_{q_{y},q_{z}}|=1$ hold after optimization, respectively. \color{black}
Optimization problem (\ref{s1b}) is still non-convex due to the rank constraint. To handle the rank constraint, we rewrite $\mathrm{C2}$ equivalently as $ 
	\widetilde{\mathrm{C2}}:\|\mathbf{W}_{m_{y},m_{z}}\|_{*}-\|\mathbf{W}_{m_{y},m_{z}}\|_{2}\ \leq 0$, see \cite{alexpower}.
Constraint $\widetilde{\mathrm{C2}}$ is in the form of a differences of two convex functions. However, $\widetilde{\mathrm{C2}}$ is still non-convex. To overcome this issue, we adopt the penalty method \cite{eaxtpenalty,alex1} and rewrite (\ref{s1b}) as follows: 
\begin{align}\label{s1b2}
&\underset  {\mathbf{W}_{m_{y},m_{z}},\alpha}{\text{maximize}}\quad\alpha- \eta^{(i)} \bigg(\|\mathbf{W}_{m_{y},m_{z}}\|_{*}-\|\mathbf{W}_{m_{y},m_{z}}\|_{2}\bigg)\\& \nonumber\text{s.t.} \; \mathrm{C1}, \mathrm{C3}, \mathrm{C4},
\end{align}
\color{black}	where $\eta^{(i)}$ is the penalty factor employed in the $i$-th iteration of the proposed \textbf{Algorithm} 1, see below. \color{black} 
The objective function in (\ref{s1b2}) is still non-convex. To tackle this issue, we determine the first-order Taylor approximation of $\|\mathbf{W}_{m_{y},m_{z}}\|_{2}$ at initial point $\mathbf{W}^{(i)}_{m_{y},m_{z}}$ as follows: 
\begin{align}& \label{appro}
\|\mathbf{W}_{m_{y},m_{z}}\|_{2}\ \geq \|\mathbf{W}^{(i)}_{m_{y},m_{z}}\|_{2}\nonumber\\& +\Tr\big[\boldsymbol{\lambda}_{\text{max}}(\mathbf{W}^{(i)}_{m_{y},m_{z}})\times\nonumber\\&\hspace{2cm} \boldsymbol{\lambda}^{H}_{\text{max}}(\mathbf{W}^{(i)}_{m_{y},m_{z}}) (\mathbf{W}_{m_{y},m_{z}}-\mathbf{W}^{(i)}_{m_{y},m_{z}})\big].
\end{align}

By substituting (\ref{appro}) into (\ref{s1b2}), we obtain the following approximated optimization problem: 
	\begin{align}\label{s1d}
&\underset  {\mathbf{W}_{m_{y},m_{z}},\alpha}{\text{maximize}}\quad\alpha -\eta^{(i)} \bigg(\|\mathbf{W}_{m_{y},m_{z}}\|_{*}-\|\mathbf{W}^{(i)}_{m_{y},m_{z}}\|_{2}\nonumber\\&-\Tr\big[\boldsymbol{\lambda}_{\text{max}}(\mathbf{W}^{(i)}_{m_{y},m_{z}})\times \nonumber\\&\hspace{2cm} \boldsymbol{\lambda}^{H}_{\text{max}}(\mathbf{W}^{(i)}_{m_{y},m_{z}}) (\mathbf{W}_{m_{y},m_{z}}-\mathbf{W}^{(i)}_{m_{y},m_{z}})\big]\bigg)\\& \nonumber\text{s.t.} \; \mathrm{C1}, \mathrm{C3}, \mathrm{C4}.
\end{align}

Optimization problem (\ref{s1d}) is convex because the objective
function is concave and the constraints span a convex set.
Therefore, it can be efficiently solved by standard convex optimization solvers such as CVX \cite{cvx}. \textbf{Algorithm} 1 summarizes
the main steps for solving (\ref{s1b}) in an iterative manner, where the
solution of (\ref{s1d}) in iteration $i$ is used as the initial point for
the next iteration $i + 1$. Assuming the maximum number of iterations $I_{\text{max}}$ is chosen sufficiently large, the algorithm produces a sequence
of improved feasible solutions until convergence to a local
optimum point of problem (\ref{Op2a}), or equivalently problem (\ref{s1b}), in polynomial time \cite{eaxtpenalty}. Codeword $\mathbf{w}_{m_{y},m_{z}}$ can be recovered as  $\mathbf{w}_{m_{y},m_{z}}=\sqrt{\delta_{\text{max}}}\mathbf{v}$, where $\mathbf{v}$  is the normalized eigenvector corresponding to the maximum and only non-zero eigenvalue $\delta_{\text{max}}=\Tr(\mathbf{W}_{m_{y},m_{z}})$. \color{black} \textbf{Algorithm} 2 summarizes the main steps for obtaining the complete reflection codebook containing $M$ codewords. \color{black}
\begin{remark}
Optimization problem (9) is an SDP problem. Thus, the complexity order of \textbf{Algorithm} 1 per iteration is  $ \mathcal{O}\bigg(\big[|\hat{\mathcal{B}}_{y,m_{y}}|+|\hat{\mathcal{B}}_{z,m_{z}}|\big]Q^{3}+\big[|\hat{\mathcal{B}}_{y,m_{y}}|+|\hat{\mathcal{B}}_{z,m_{z}}|\big]^{2}Q^{2}+\big[|\hat{\mathcal{B}}_{y,m_{y}}|+|\hat{\mathcal{B}}_{z,m_{z}}|\big]^{3}\bigg)\approx \mathcal{O}\big(\big[|\hat{\mathcal{B}}_{y,m_{y}}|+|\hat{\mathcal{B}}_{z,m_{z}}|\big]Q^{3}\big)$, where $\mathcal{O}$ is the big-O notation. \color{black}
\end{remark}
	\begin{algorithm}[H]\label{A11}
	\begin{algorithmic}[1]
		\STATE {Initialize:} Generate random initial matrix $\mathbf{W}_{m_{y},m_{z}}^{(1)}$, set iteration index $i=1$, maximum number of iterations $I_{\text{max}}$, \color{black} $\alpha>0$, and penalty factors $\eta^{(1)}>0$ and $\eta_{\text{max}}$ \color{black} \\
		\STATE \textbf{Repeat}\\
		\STATE Solve convex problem (\ref{s1d}) for given   $\mathbf{W}_{m_{y},m_{z}}^{(i)}$, and store the intermediate solution   $\mathbf{W}_{m_{y},m_{z}}$\\
		\STATE Set ${i}={i}+1$ and update  $\mathbf{W}_{m_{y},m_{z}}^{(i)}=\mathbf{W}_{m_{y},m_{z}}$ and \color{black} $\eta^{(i)}=\min(\alpha \eta^{(i-1)},\eta_{\text{max}})$ \color{black} 
		\STATE \textbf{Until} convergence or $i=I_{\text{max}}$ \\		
		\STATE {Output:} $\mathbf{w}_{m_{y},m_{z}}^{*}=\mathbf{w}_{m_{y},m_{z}}$
	\end{algorithmic}
	\caption{Successive Convex Approximation (SCA)}
\end{algorithm}\vspace{-0.5cm}
\begin{algorithm}[H]
	\begin{algorithmic}[1]
		\STATE {Input:} The number of IRS elements $Q$ and the collection of intervals $ \mathcal{B}_{y,m_{y}}$ and $ \mathcal{B}_{z,m_{z}}$, i.e., codebook size $M$ 
				\STATE \textbf{for} $m_{y}=1:M_{y}$ 
		\STATE  \textbf{for} $m_{z}=1:M_{z}$ 
		\STATE $\;\;$Use \textbf{Algorithm} 1 to design codeword $\mathbf{w}_{m_{y},m_{z}}$
		\STATE \textbf{end for} 
		\STATE \textbf{end for} 
		\STATE {Output:} The designed codewords $\mathbf{w}_{m_{y},m_{z}}, \forall m_{y}=\{1,\dots,M_{y}\}, \forall m_{z}=\{1,\dots,M_{z}\}$
	\end{algorithmic} \label{A2}
	\caption{Reflection Codebook Generation}
\end{algorithm} \vspace{-0.5cm}
\section{Discrete Phase-Shift Codebook Design}\color{black}
For ease of practical implementation, we also
consider the case where the phase-shift of each IRS element can take only a finite number of discrete values which are collected in set $
\mathcal{S}=\{0, \Delta \nu, \dots, (S-1)\Delta \nu \}$, 
where $\Delta \nu=\frac{2\pi}{S}$, $S=2^{b}$, and $b$ denotes the number of bits needed to specify the $S$ phase-shift levels.   

\subsection{Optimization Problem}
Similar to problem (\ref{Op2a}) for continuous phase shifts, the discrete phase-shift optimization problem can be formulated as follows:
\begin{align}\label{OP2}
&\underset  {\mathbf{w}_{m_{y},m_{z}},\alpha}{\text{maximize}}\quad\alpha \\& \nonumber\text{s.t.} \; \mathrm{C1}:|g_{m_{y},m_{z}}(\beta_y,\beta_z)|^{2} =|\mathbf{w}_{m_{y},m_{z}}^{H} (\by (\beta_{y}) \otimes \bz (\beta_{z}))|^{2}  \geq \alpha , \nonumber\\&\hspace{3cm} \forall \beta_y \in  \hat{\mathcal{B}}_{y,m_{y}}, \beta_z \in  \hat{\mathcal{B}}_{z,m_{z}}, \nonumber\\& \;\quad
\mathrm{C2}: \nu_{q} \in \bigl\{\mathcal{S}\bigl\}, \forall q= \{0, \dots, Q-1\}.	\nonumber
\end{align}
 Optimization problem (\ref{OP2}) is a non-convex mixed-integer problem which is difficult to solve. However, we show in the following that problem (\ref{OP2}) can be transformed into the canonical form of a binary linear problem which allows the application of CVX to solve it optimally. \color{black}
\subsection{Proposed Solution}
We start by writing
\begin{align}&
|\mathbf{w}_{m_{y},m_{z}}^{H} (\by (\beta_{y}) \otimes \bz (\beta_{z}))|^{2}=\mathbf{w}_{m_{y},m_{z}}^{H} \mathbf{A}_{\beta_{y},\beta_{z}}\mathbf{w}_{m_{y},m_{z}},	
\end{align}
where $\mathbf{A}_{\beta_{y},\beta_{z}}=\big(\by (\beta_{y}) \otimes \bz (\beta_{z})\big) \big(\by (\beta_{y}) \otimes \bz (\beta_{z})\big)^{H}$ is a positive semi-definite matrix.
We can further rewrite $\mathbf{w}_{m_{y},m_{z}}^{H} \mathbf{A}_{\beta_{y},\beta_{z}}\mathbf{w}_{m_{y},m_{z}}$ as follows:
\begin{align}&\label{opd}
\mathbf{w}_{m_{y},m_{z}}^{H} \mathbf{A}_{\beta_{y},\beta_{z}}\mathbf{w}_{m_{y},m_{z}} 
= \sum_{q=0}^{Q-1}\sum_{i=0}^{Q-1}|\mathbf{A}_{\beta_{y},\beta_{z}}^{(q,i)}|\big(\cos(\text{Ang}(\mathbf{A}_{\beta_{y},\beta_{z}}^{(q,i)}))\\&+j\sin(\text{Ang}(\mathbf{A}_{\beta_{y},\beta_{z}}^{(q,i)}))\big)\bigg(\cos(\nu_{q}-\nu_{i})+j\sin(\nu_{q}-\nu_{i})\bigg),\nonumber \\&=\sum_{q=0}^{Q-1}\sum_{i=0}^{Q-1}|\mathbf{A}_{\beta_{y},\beta_{z}}^{(q,i)}|\bigg(\cos(\nu_{q}-\nu_{i})\cos(\text{Ang}(\mathbf{A}_{\beta_{y},\beta_{z}}^{(q,i)}))\nonumber\\&-\sin(\text{Ang}(\mathbf{A}_{\beta_{y},\beta_{z}}^{(q,i)}))\sin(\nu_{q}-\nu_{i})\bigg),\nonumber
\end{align}
where $\mathbf{A}_{\beta_{y},\beta_{z}}^{(q,i)}$ is the $(q,i)$-th entry of matrix $\mathbf{A}_{\beta_{y},\beta_{z}}$, and $\text{Ang}(\mathbf{A}_{\beta_{y},\beta_{z}}^{(q,i)})$ is the phase of $\mathbf{A}_{\beta_{y},\beta_{z}}^{(q,i)}$. Optimization problem (\ref{opd}) is non-linear with respect to the phase-shift due to the non-linear functions $\cos(\cdot)$ and $\sin(\cdot)$. To transform these non-linear functions, we apply a linear transformation as in \cite{irsdiscrete}, \cite{discreteirs2}. Let us define $
\nu_{q}=\bar{\mathbf{a}}^{T}\mathbf{x}_{q},
$ 
where $\mathbf{x}_{q}$ is a binary vector of dimension $2^{b}$ satisfying $\|\mathbf{x}_{q}\|=1$, and 
$
\bar{\mathbf{a}}=\big[0, \Delta \nu, \dots, (S-1)\Delta \nu  \big]
$.

Moreover, we define $\Delta \nu_{q,i}=\nu_{q}-\nu_{i}$. Similarly, we define binary vector $\mathbf{y}_{q,i}$ of dimension $2^{b+1}-1$. Thus,  the following relation holds:
$
\bar{\mathbf{a}}^{T}(\mathbf{x}_{q}-\mathbf{x}_{i})=\mathbf{a}^{T}\mathbf{y}_{q,i}$, 
where 
$
\mathbf{a}=\big[-(S-1)\Delta \nu,\dots,-\Delta \nu,0, \Delta \nu, \dots, (S-1)\Delta \nu  \big],
$ 

Optimization problem (\ref{OP2}) can be rewritten as the following equivalent mixed binary linear optimization problem: 
\begin{align}\label{opfd}
&\underset  {\mathbf{x},\mathbf{y},\alpha}{\text{maximize}}\quad\alpha \\& \text{s.t.} \; \mathrm{C1}: \sum_{q=0}^{Q-1}\sum_{i=0}^{Q-1}|\mathbf{A}_{\beta_{y},\beta_{z}}^{(q,i)}|\bigg(\cos(\text{Ang}(\mathbf{A}_{\beta_{y},\beta_{z}}^{(q,i)}))\mathbf{c}^{T}\nonumber\\&-\sin(\text{Ang}(\mathbf{A}_{\beta_{y},\beta_{z}}^{(q,i)}))\mathbf{s}^{T}\bigg)\mathbf{y}_{q,i} \geq \alpha,  \forall \beta_y \in  \hat{\mathcal{B}}_{y,m_{y}}, \beta_z \in  \hat{\mathcal{B}}_{z,m_{z}}, \nonumber\\& 
\mbox {C2}: x_{q}(j)\in \{0,1\}, \forall j, q,\;\; \mathrm{C3}:\sum_{j=1}^{2^{b}}x_{q}(j)=1, \forall q, \nonumber\\& \mbox {C4}: y_{q,i}(u) \in \{0,1\}, \forall u, q,i,\nonumber \nonumber\\& \mbox {C5}: \sum_{u=1}^{2^{b+1}-1}{y}_{q,i}(u)=1, \forall q,i,\nonumber\mbox {C6}: \bar{\mathbf{a}}^{T}(\mathbf{x}_{q}-\mathbf{x}_{i})=\mathbf{a}^{T}\mathbf{y}_{q,i}, \forall  q,i.\nonumber
\end{align}
where $\mathbf{y}$ denotes the collection of optimization variables $\mathbf{y}_{q,i}, \forall q,i,$ $x_{q}(j)$ is the $j$-th entry of vector $\mathbf{x}_{q}$, $y_{q,i}(u)$ is the $u$-th entry of vector $\mathbf{y}_{q,i}$, $
\mathbf{c}^{T}=\cos(\mathbf{a}^{T}), $ and $
\mathbf{s}^{T}=\sin(\mathbf{a}^{T})$. Optimization problem (\ref{opfd}) is in the form of a binary linear optimization problem which can be solved using CVX \cite{cvx}. The phase of the $q$-th element of $\mathbf{w}_{m_{y},m_{z}}$ is given by $\bar{\mathbf{a}}^{T}\mathbf{x}_{q}$. \color{black}
\color{black}
\begin{remark}
The complexity of problem (\ref{opfd}) is exponential
in $Q$ and $|\mathcal{S}|$. However, (\ref{opfd}) can be solved efficiently using CVX\cite{cvx}.   
\end{remark}
\color{black}
\begin{remark}
	The discrete phase-shift reflection codebook can be generated using \textbf{Algorithm} 2 by solving problem (\ref{opfd}) to design the codewords $\mathbf{w}_{m_{y},m_{z}}, \forall m_{y}, m_{z}$.  
\end{remark}
\section{Simulation Results}

	\begin{figure}
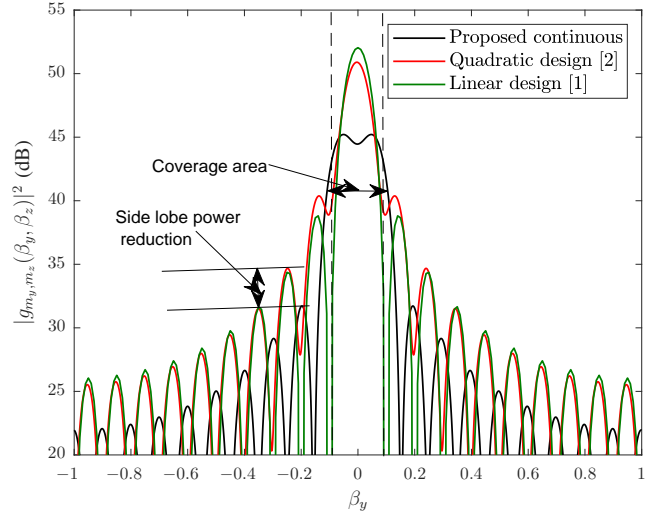
\vspace{-0.25cm}
		\centering
		\scalebox{0.65}{
			\psfragfig{fig1a}{
		}}
		\caption{$|g_{m_{y},m_{z}}(\beta_y,\beta_z)|^{2}$ in \textrm{dB} vs. $\beta_{y}$ for different codebook designs, $Q=400$, $I_{\textrm{max}}=25$, $d_{y}=d_{z}=\frac{\lambda}{2}$.}
		\label{Fig1}\vspace{-0.45cm}
	\end{figure}
	\hspace*{0.1cm}
	\begin{figure}
		\centering
		\scalebox{0.65}{
			\psfragfig{fig4}{}}
		\caption{$|g_{m_{y},m_{z}}(\beta_y,\beta_z)|^{2}$ in \textrm{dB} vs. $\beta_{y}$ for continuous and discrete phase-shift designs, $I_{\textrm{max}}=25$, $d_{y}=d_{z}=\frac{\lambda}{2}$, $Q=100$.}
		\label{Fig4}\vspace{-0.45cm}
	\end{figure}
	\hspace*{0.1cm}
	\begin{figure}
		\centering
		\scalebox{0.65}{\psfragfig{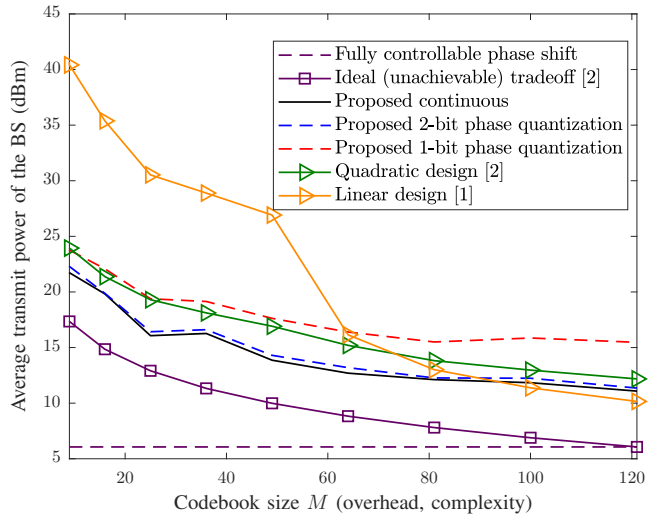}}
		\caption{Average total transmitted power versus codebook size, $Q=121$, $\gamma_{\text{req}}=10$~dB.}
		\label{Fig4f}\vspace{-0.25cm}
	\end{figure} \vspace{-0.25cm}

\begin{figure*}[h]	
	\begin{subfigure}{0.24\textwidth}
		\centering
		\scalebox{0.24}{
			\psfragfig{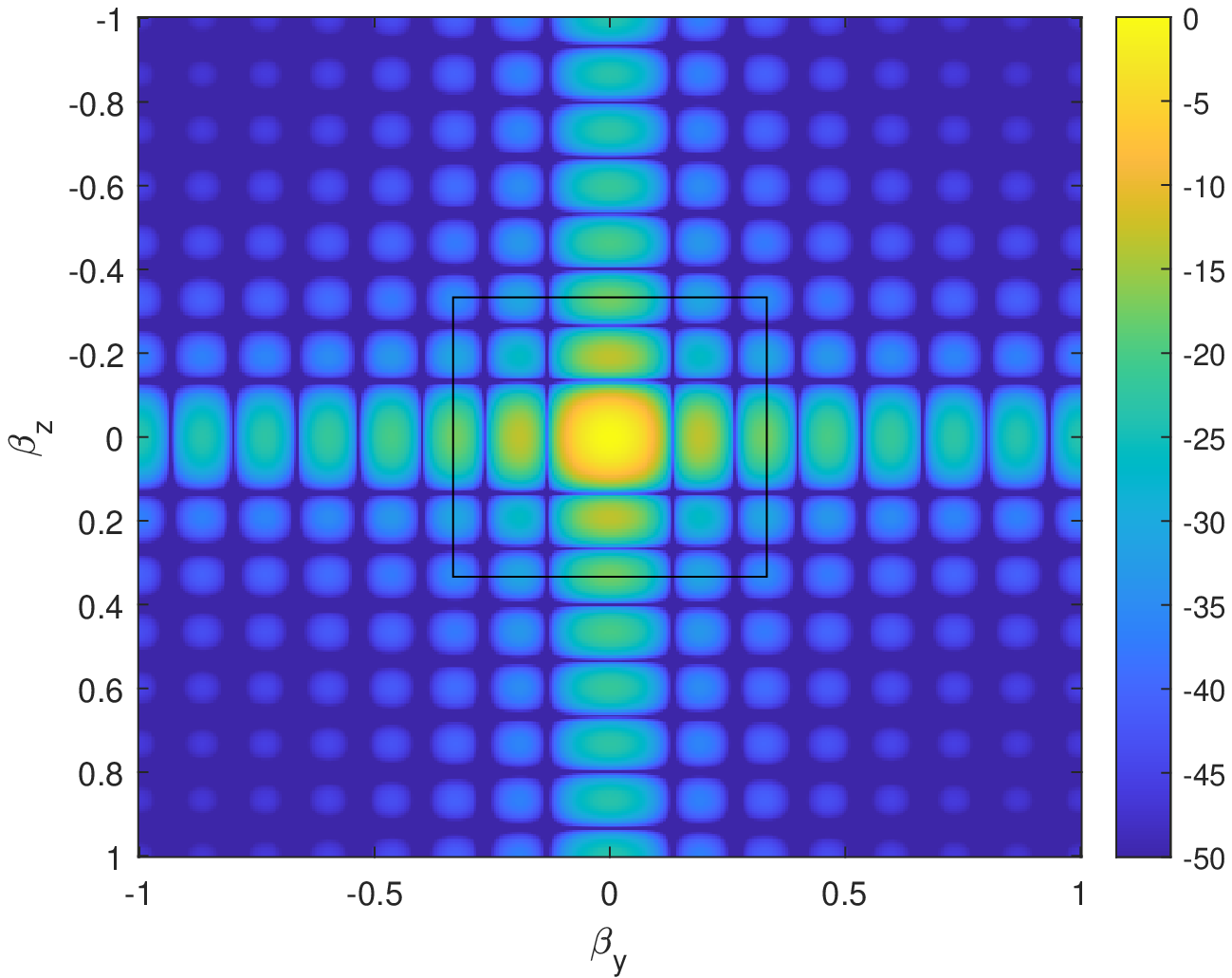}{
		}}
		\caption{Linear design}
		\label{Figh1}
	\end{subfigure}
	\begin{subfigure}{0.24\textwidth}
		\centering
		\scalebox{0.24}{
			\psfragfig{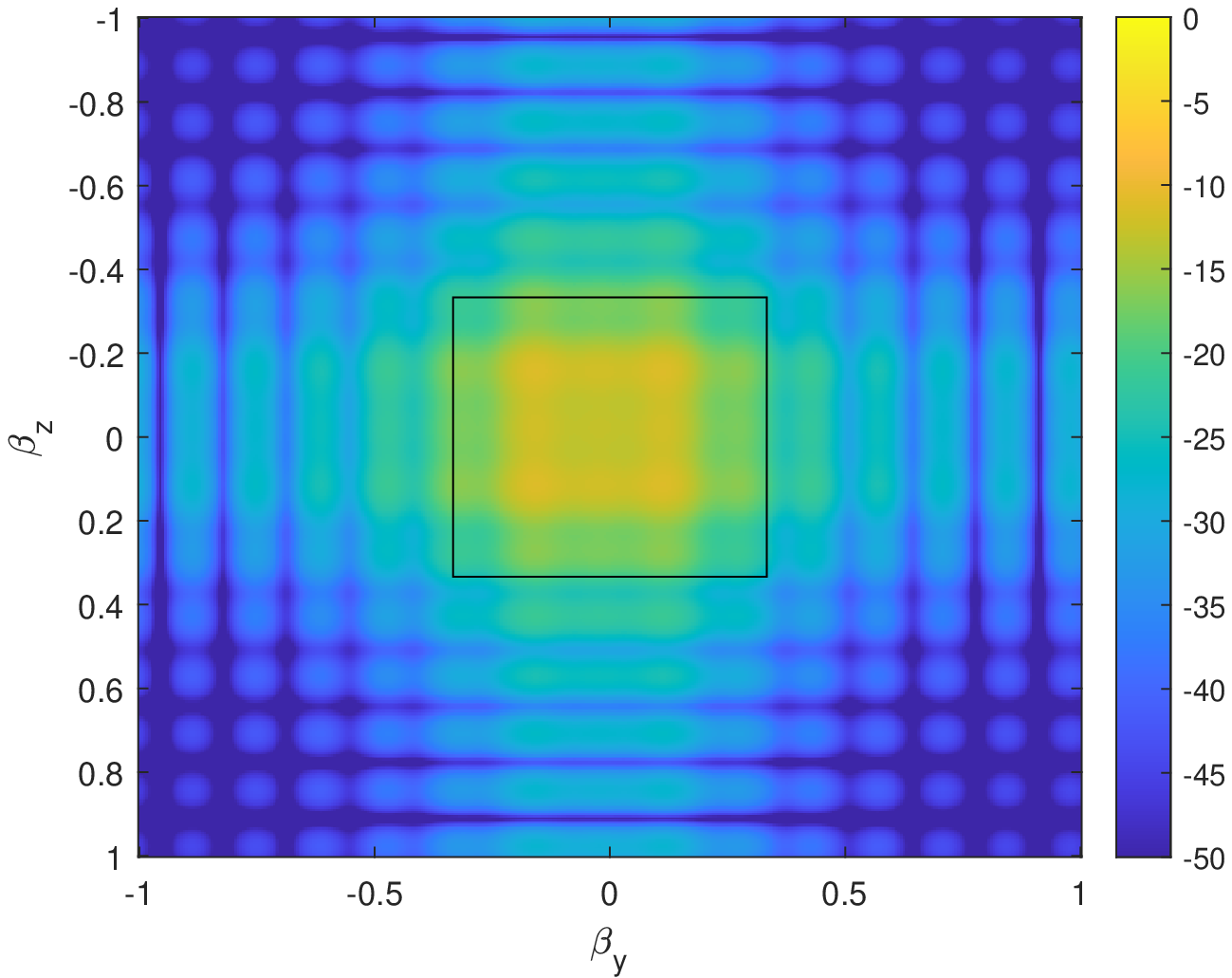}{
		}}
		\caption{Quadratic design}
		\label{Figh2}
	\end{subfigure}
	\begin{subfigure}{0.24\textwidth}
		\centering
		\scalebox{0.24}{
			\psfragfig{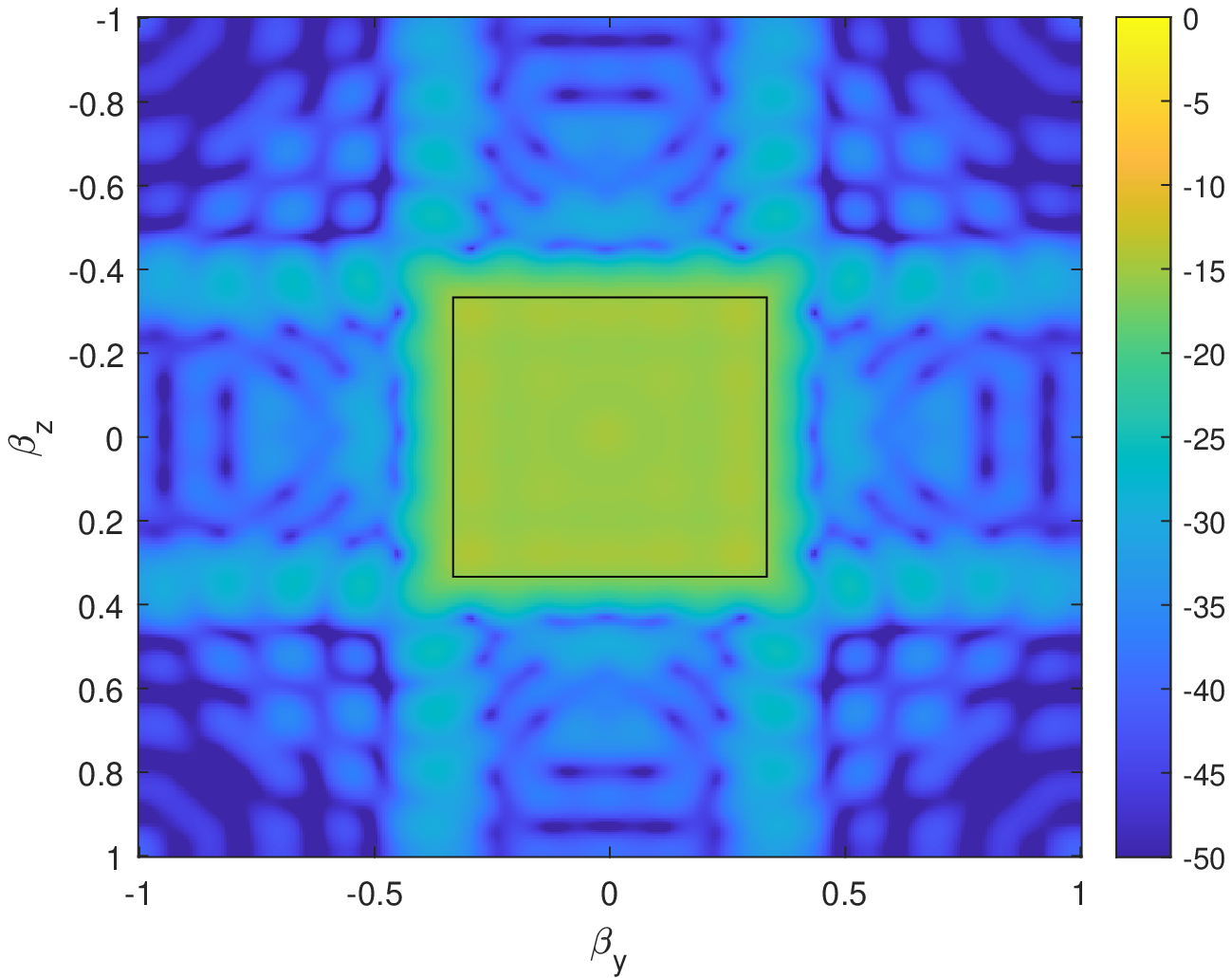}{
		}}
		\caption{Continuous design}
		\label{Figh3}
	\end{subfigure}	
	\begin{subfigure}{0.24\textwidth}
		\centering
		\scalebox{0.24}{
			\psfragfig{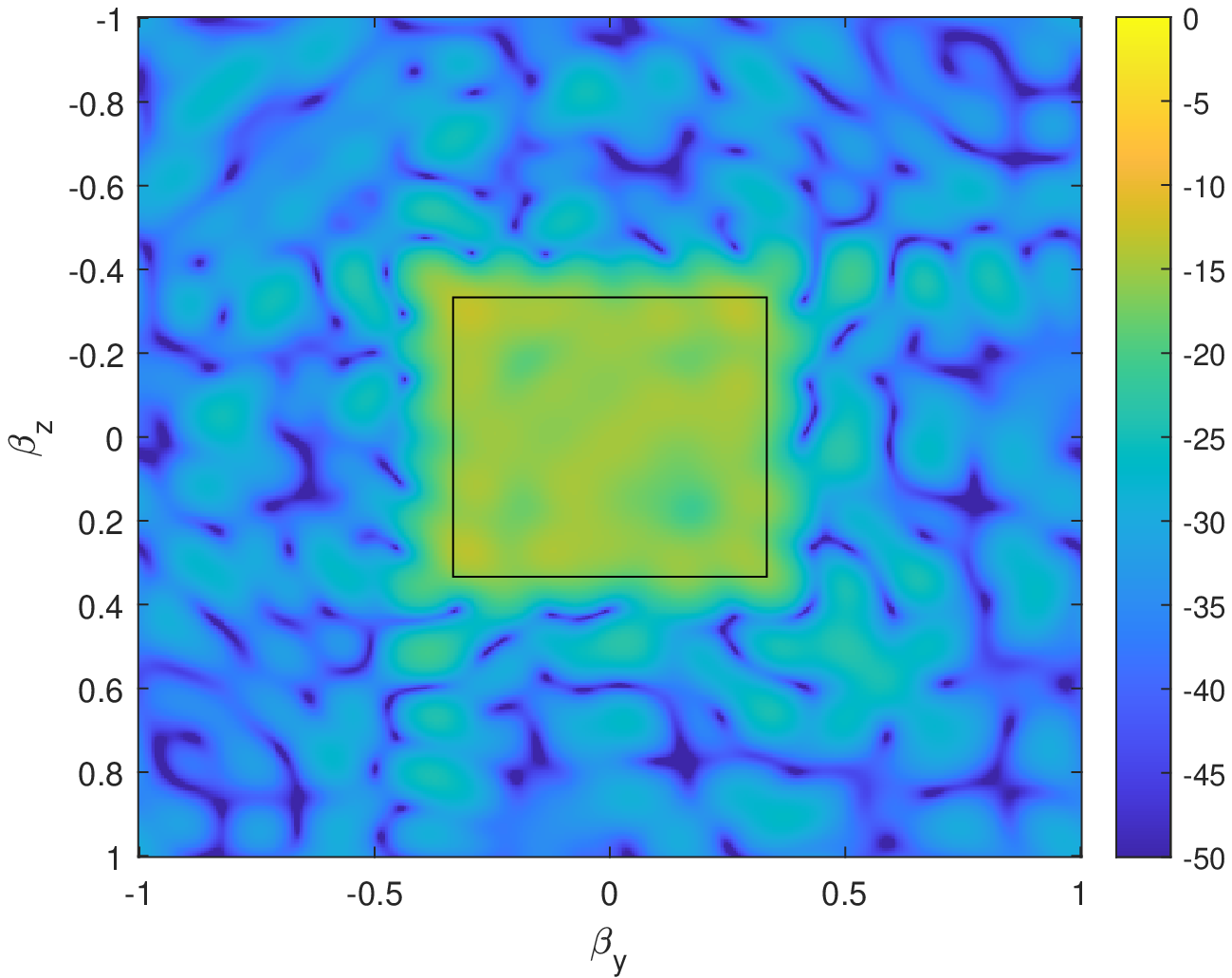}{
		}}
		\caption{2-bit phase quant.}
		\label{Figh4}
	\end{subfigure}	
		\caption{Heat maps for different codebook designs, $Q=225$, $M=9$, $I_{\textrm{max}}=25$, $d_{y}=d_{z}=\frac{\lambda}{2}$.}\vspace{-0.5cm}
\end{figure*}\vspace{-0.25cm}

For simplicity of presentation and comparison, unless specified otherwise, we focus on the phase variation along the $y$-axis (i.e., $\beta_{z}=0$). In Fig. \ref{Fig1}, we show $|g_{m_{y},m_{z}}(\beta_y,\beta_z)|^{2} $ in \textrm{dB} vs. $\beta_{y}$ for different codebook designs, namely, the proposed optimized continuous design, the linear design \cite{marirsj}, and the quadratic design \cite{quad}. We consider a codebook of size $M=169$ and design intervals $ \mathcal{B}_{y,m_{y}}=[-0.0769~0.0769]$ and $ \mathcal{B}_{z,m_{z}}=[-0.0769~0.0769]$. The remaining system parameters are given in the caption of the figure. 
As can be seen from Fig. \ref{Fig1}, the proposed continuous design generates a flat beam within $ \mathcal{B}_{y,m_{y}}$, which is generally desirable for uniform coverage. On the other hand, the linear design fails to achieve uniform coverage within $ \mathcal{B}_{y,m_{y}}$, as the reflected power is large in the center of the interval while it is very small at the edge of the interval. \color{black} Compared with the quadratic design, the proposed design yields a flatter beam in the coverage area. Furthermore, for the proposed design, the side lobes are much smaller compared to the linear and quadratic designs.  \color{black}  

Fig. \ref{Fig4} shows the impact of the number of quantization levels on the shape of the beam reflected from the IRS, i.e., $|g_{m_{y},m_{z}}(\beta_y,\beta_z)|^{2} $. We aim to maximize the power reflected from the IRS for a codebook size of $M=81$ and consider the intervals of $ \mathcal{B}_{y,m_y{}}=[-0.11~0.11]$ and $ \mathcal{B}_{z,m_{z}}=[-0.11~0.11]$. We observe that for a $1-$bit
phase-shift quantization, the main lobe of the beam is reduced and the
side lobes deviate from those for ideal non-quantized phase shifts. On the other hand, for $b\geq 2$, the main lobe is close to that for the continuous phase-shift case.

In Fig. \ref{Fig4f}, we show the trade-off between the average transmit power of the BS required to achieve an SNR of $\gamma_{\text{req}}=10$~dB at the user and the codebook size $M$. We assume the direct link between the BS and the user is blocked. We consider LoS channels between the BS and the IRS and between the IRS and the user. The corresponding link distances are $d_{1}=d_{2}=20$ \textrm{m}. The free-space path loss of the links is $P_{Lt}=\big(c/(4\pi d_{1}f)\big)^{2}$ and $P_{Lr}=\big(c/(4\pi d_{2}f)\big)^{2}$, where $c$ is the speed of light and $f=3.4~\textrm{GHz}$ is the carrier frequency. The bandwidth is $20$ \textrm{MHz} and the noise power spectral density is $-174$~dBm/Hz. The results in Fig. \ref{Fig4f} were obtained by averaging over $10^{5}$ uniformly random realizations of the AoAs and AoDs of the IRS. The transmit power required to ensure $\gamma_{\text{req}}$ is given by $
p_{\text{req}}=\frac{\gamma_{\text{req}}\sigma^{2}}{\bar{g}^{2}\max_{\mathbf{w}_{m_{y},m_{z}}}\|\mathbf{h}^{H}{\text{Diag}}{(\mathbf{w}_{m_{y},m_{z}})}\mathbf{g}\|^{2}}, 
$

where $\sigma^{2}$ is the receiver noise power, $\mathbf{h}\in \mathbb{C}^{Q}$ is the channel vector between IRS and user, and $\mathbf{g}\in \mathbb{C}^{Q}$ is the channel vector between BS and IRS.  $\mathbf{h}$ and $\mathbf{g}$ are modelled using the path losses and the steering vectors at the IRS. \color{black} We also show two performance upper bounds, one achieved with fully-controllable IRS phase shifts (asymptotically achievable as $M\to\infty$) and an unattainable bound for $M\leq Q$ given in [2] leading to required transmits powers of $
p_{\text{req},F}=\frac{\gamma_{\text{req}}\sigma^{2}}{\bar{g}^{2}P_{Lt}P_{Lr}Q^{2}} 
$ and $p_{\text{req},I}=\frac{\gamma_{\text{req}}\sigma^{2}}{\bar{g}^{2}P_{Lt}P_{Lr}QM}$, respectively. \color{black}  

As can be seen from Fig.~\ref{Fig4f}, the proposed codebook designs require a lower transmit power than the linear and the quadratic design, especially for small codebook sizes, $M \ll Q $, which is the regime of interest in practice. This is due to the fact that the proposed designs cover the design interval of $ \mathcal{B}_{y,m_{y}}$ more uniformly than the baseline schemes. For $M\approx Q$, the linear design slightly outperforms the proposed and the quadratic designs. This is because although the linear design yields the most narrow beam, with $M=Q$, it can still cover the entire angular space. Hence, further widening of the beam, as is done for the proposed and the quadratic designs, is not required and degrades the performance.

Fig. \ref{Fig4f} also shows the impact of phase-shift quantization on the average total transmit power. We observe that a $1-$bit phase-shift quantization requires a higher average transmitted power compared to continuous phase shifts. This is expected since due to coarse discrete phase shifts, the signals reflected by the IRS cannot be perfectly aligned in phase at the receiver, thus resulting in a power loss. In contrast, for $2-$bit quantization, the performance is close to that for continuous phase shifts.
 
\color{black}
Figs.~\ref{Figh1}, \ref{Figh2}, \ref{Figh3}, and \ref{Figh4} show 2D heat-map plots for $|\frac{\bar{g}}{g_{\text{max}}}g_{m_{y},m_{z}}(\beta_y,\beta_z)|^{2}$ in \textrm{dB}, where $g_{\text{max}}=\bar{g}Q$, see (\ref{res}), for different codebook designs. The IRS size is $Q=225$. We consider $M=9$ and design intervals $ \mathcal{B}_{y,m_{y}}=[-0.3333 ~0.3333 ]$ and $ \mathcal{B}_{z,m_{z}}=[-0.3333 ~0.3333 ]$. As can be seen, the proposed continuous design attains a more uniform coverage in the area of interest compared to the linear and quadratic designs. Moreover, Fig.~\ref{Figh4} reveals that for 2-bit quantization, the proposed algorithm also provides uniform coverage in the area of the main lobe.\color{black}

\vspace*{-0.2cm}
\section{Conclusions}\vspace*{-0.25cm}
In this paper, we proposed a new codebook construction method to obtain a set of pre-designed phase-shift configurations for IRS unit cells. We considered both continuous and discrete phase-shift designs and formulated the codebook construction as optimization problems. We proposed an optimal algorithm for the discrete phase-shift design and a low-complexity sub-optimal solution for the continuous phase-shift design. Simulation results illustrated that the proposed algorithms support different codebook sizes with different beamwidths and they enable significant power savings compared to the existing linear and quadratic codebook designs \cite{marirsj,quad}, especially for practical small codebooks. The proposed codebook designs can be used for beam training and online optimization for IRS-assisted wireless communication systems. 
\bibliography{ref}  
\bibliographystyle{IEEEtran}
\end{document}